# Long-run in-operando NMR to investigate the evolution and degradation of battery cells

Steffen A. Kayser[†§], Achim Mester[‖], Andreas Mertens[†], Peter Jakes[†], Rüdiger-A. Eichel[†#], Josef Granwehr[†§]

[†]*Forschungszentrum Jülich, Institute of Energy and Climate Research – Fundamental Electrochemistry (IEK-9), 52425 Jülich, Germany.*

[§]*RWTH Aachen University, Institute of Technical and Macromolecular Chemistry, 52074 Aachen, Germany.*

[‖]*Forschungszentrum Jülich, Central Institute of Engineering – Electronics and Analytics – Electronic Systems (ZEA-2), 52425 Jülich, Germany.*

[#]*RWTH Aachen University, Institute of Physical Chemistry, 52074 Aachen, Germany.*

**ABSTRACT**

Nuclear magnetic resonance (NMR) investigations of electrochemical systems require gas-tight and non-metallic cell housings. This contribution reports on the development and evaluation of a cylindrical battery container in combination with a numerically optimized saddle coil that is suitable for NMR investigations of battery cells over hundreds of charge–discharge cycles. The reliability of the new cell container design and its long-time gas-tight sealing are shown by rate capability comparisons to standard housings with $LiCoO_2$ (LCO) vs. Li-metal electrodes as well as a charge–discharge experiment of a LCO vs. graphite batteries over more than 2000 hours. To demonstrate the performance of the entire NMR setup, long-run in-operando measurements on a Li-metal vs. graphite cell are presented. The NMR data reveal the formation and evolution of mossy and dendritic Li microstructures over a period of 1000 h. Analyzing the measured rate of microstructure growth could help to identify dendrite mitigation strategies, such as enhanced cell pressure or additives, and could enable a method for battery lifetime prediction.

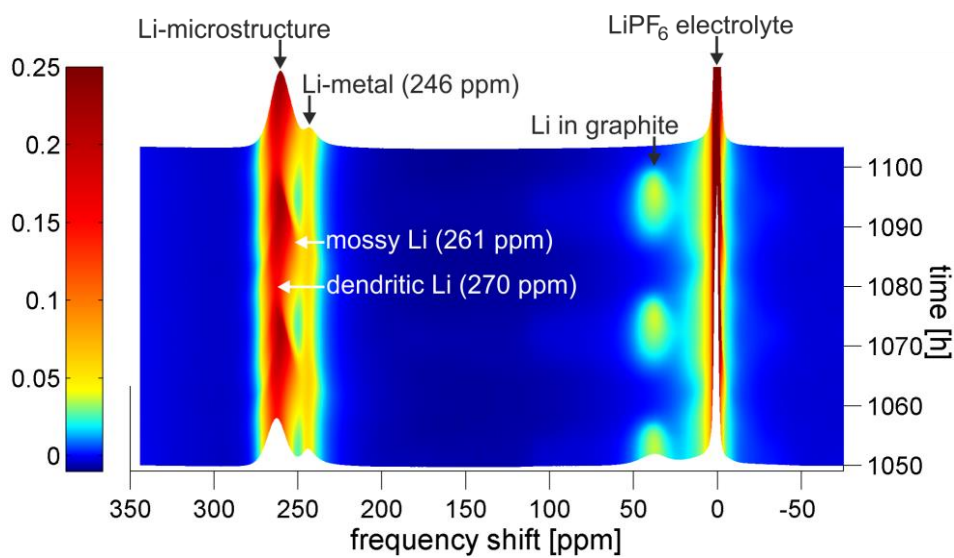

# 1. INTRODUCTION

To develop and improve state-of-the-art battery systems, in-operando nuclear magnetic resonance (NMR) presents a viable option to reveal transport properties and transformation reactions on microscopic as well as on macroscopic length scales[1–4]. The first in-situ experiments on lithium-ion batteries (LIBs) were performed using coin and cylindrical cell containers with integrated NMR coils[5–7]. For coin cells with metal housings, a solid copper disk inside these batteries acted simultaneously as current collector and as NMR radio frequency (rf) central conductor for excitation and detection. In the cylindrical cell design a straight centered wire was used for excitation/detection and current collection. For these in-situ NMR measurements the battery operation had to be stopped, but the cell was charged and discharged inside the NMR magnet.

The first in-operando NMR measurements were performed using a solenoid coil wherein a hermetically sealed bag containing a LIB was placed[8]. The cell was connected to a battery cycler and the NMR spectra were acquired during charging and discharging. One possible design suitable for bag cells is Bellcore's laminated cell[9], which is enclosed in heat-sealable plastic bags[10]. For the electrical connections, meshes attached to the electrodes have been generally used[11–17]. Conductive foils for the connections are possible as well, but sealing is known to be more challenging[18–20]. Bag cells are flexible and low-cost. They do not require external pressure to maintain contact, but undefined or missing pressure on the cell components can lead to increased resistance inside the battery and poor reproducibility. Because most bag materials without coating are permeable to air, the lifetime of a bag cell is rather short. For a LIB in a polyester bag a lifetime of around five days has been reported[21]. By using aluminum-coated bags, which protect the cells for a longer time but decrease the overall NMR signal intensity, the capacity fading of LIBs within 300 cycles was investigated[22].

Another promising approach for in-operando measurements on batteries and other electrochemical systems are cylindrical cell containers[23]. Inside the NMR magnet a horizontal or a vertical orientation is possible. The orientation of the electrodes is particularly important with regard to bulk susceptibility effects[14,21,24] and rf shielding[25]. The NMR signal of a Li-metal electrode is shifted by approximately 245 ppm in case of an orientation perpendicular to the external static magnetic field, $B_0$, and shifted by around 272 ppm in case of a parallel orientation. This can be attributed to the temperature independent paramagnetism of Li-metal[21]. An electrode orientation perpendicular to the $B_1$ field leads to weak signals from the main faces of the electrodes but high signals from the edges. Using an orientation parallel to $B_1$ enables an almost uniform excitation of the electrode's main faces[24]. When the cell container is placed lying in a solenoid rf coil, disk-shaped electrodes can be oriented parallel to the $B_0$ field[23,26]. By using specifically designed inserts[27] for the cell assembly the electrode orientation can be turned to enable a horizontal arrangement in a solenoid rf coil[28–30]. Configurations with electrodes perpendicular to $B_0$ and parallel to $B_1$ can be realized as well by using Helmholz, saddle, or Alderman–Grant rf coils placed around a vertically standing cylindrical cell container[6,31,32].

Numerical simulations offer great potential for the optimization of rf coils in NMR experiments. Thereby calculation of the electrode orientation dependent rf shielding and $B_1$ field strength becomes possible[24]. At present, the most common approach is the finite-element method (FEM) that offers a broad flexibility in the design of the electrical system and enables the analysis of effects due to dielectric materials in the vicinity of the coils[33,34]. An alternative method to optimize the impedance of an rf coil design is the method of moments (MoM) that is used in software packages such as NEC[35]. This technique, however, has been rarely used for magnetic resonance probe design to date.

The filling factor and thereby the NMR signal intensity can be enhanced by increasing the electrode separator thickness. A cylindrical container design with approximately 8 mm spacing between the electrodes was used to study the correlation between the electrolyte concentration gradient and microstructure growth on a Li-metal electrode with one-dimensional magnetic resonance imaging (MRI)[36]. Similar cylindrical setups were tested by other groups before[37,38]. Such vertically standing battery housings were used for slice selective probing of a LIB in the stray field of a permanent magnet[39,40] and in combination with an additional coil for pulsed field gradients[31,32,41]. Moreover, three-dimensional $^1$H MRI was applied to investigate the Li microstructure growth indirectly[42]. Crucial for all types of cell containers are sealing systems, current collector design and the choice of chemically resistant materials. For cylindrical cells, wires and screws were tested as well as sealing by rings or epoxy[23,31,40,43].

Batteries containing metallic Li exhibit a very high energy density (3860 mAh/g theoretical specific capacity), yet dendrite growth prevents large cycle numbers and pose an inherent safety risk. Therefore understanding the formation of mossy and dendritic Li structures is of rising interest[12,13,16,29,44]. To our knowledge, long-time in-operando NMR experiments of the Li microstructure growth have not been reported so far.

In a Li-metal vs. graphite battery Li gets reversibly intercalated into the graphite electrode during discharge and deintercalated during charge[45]. Using $^7$Li in-operando NMR the formation of different stages corresponding to $LiC_{18}$, $LiC_{12}$ and $LiC_6$ leads to spectral lines at 13 ppm, 42 ppm and 37 ppm, respectively[18,19,22,46–48]. During charge the Li is deposited in a microstructured morphology on the Li-metal electrode[44,49,50]. The morphology of this Li microstructure depends on various parameters like temperature, charge/discharge rate and the electrolyte composition[51,52]. The growth of these microstructures can be measured by in-operando NMR as well. Their signals overlap with the Li-metal signal and with each other, but the line centers differ. A fitting of the spectrum can be used to distinguish between mossy (261 ppm), dendritic (270 ppm) and metallic (246 ppm) structures. This was verified by simulations, scanning electron microscopy (SEM) and $^6$Li/$^7$Li isotope NMR studies[16]. Bulk magnetic susceptibility effects lead to orientation dependent frequency shifts[24,29]. It is assumed that mossy Li grows parallel to the electrode surface and more densely then dendritic Li, which is assumed to grow more or less perpendicularly to the surface, thus parallel to the direction of the applied electric field[53].

In the following a gas tight NMR cell container design is described, which enables long-time investigations. This is demonstrated by current rate capability tests, a long-time (2400 h) cycling experiment and electrical impedance spectroscopy (EIS). Moreover the possibility for $^7$Li in-operando experiments on a Li-metal vs. graphite battery before and after a working time of 1100 h is demonstrated. Thereby increasing mossy and dendritic Li microstructures with charge–discharge dependent growth and depletion are described.

## 2. EXPERIMENTAL AND COMPUTATIONAL METHODS

### 2.1 Cell container design

The in-operando cell container is shown in Fig. 1. In total it is 93 mm long and has an outer diameter of 15 mm at the position of the electrochemical cell components. Conceptually, its body is a cylinder with a drilled hole. The battery cell is positioned at the bottom of this 12.5 mm inner diameter bore and closed with a shaft. The shaft is pushed into the body and then fixed by a screw nut, which presses a cutting ring. The cutting ring, which is a kind of surface seal O-ring, has a triangular cross section and seals the container reliably (overpressure up to 15 bar was tested; the $N_2$ pressure loss was below 1 bar within one day starting at 15 bar). The force for compressing cell components can be applied reproducibly without any turning motion between body and shaft. The spacing between the electrodes can be adjusted with a tolerance below 0.1 mm.

For electrical contacting of the cell, contacting sticks are pushed through holes in the body and the shaft. A diameter reduction of these sticks supports a correct alignment at end stops inside body and shaft. They cannot be pressed deeper than 0.1 mm into the cell chamber to protect the cell components as well as for a reliable electrical contact. The contacting sticks are sealed by cutting rings as well. By completely avoiding epoxy or glue, a contamination or absorption of the electrolyte is prevented. Moreover the container can be opened and reassembled many times.

Contacting sticks can be easily substituted, as they are not permanently incorporated into the cell housing. For long-run experiments on LIBs with metallic Li electrodes, solid copper contacting sticks were used that did not show a contact loss due to slow chemical reactions of the copper with Li-metal to form Li-Cu alloys[54]. In addition to full-metallic sticks, plastic sticks with thin conductive coatings have been used. Thereby less metal is loaded into the sensitive region of the rf coil, leading to a reduction of damping, distortions and shielding of the rf magnetic field, $B_1$. Contacting layers were applied by cold-plasma sputtering (Bal-Tec SCD 050). For the presented experiments, copper was used. A layer thickness of 500 nm is sufficient for low direct current (DC) resistance and adequate layer stability.

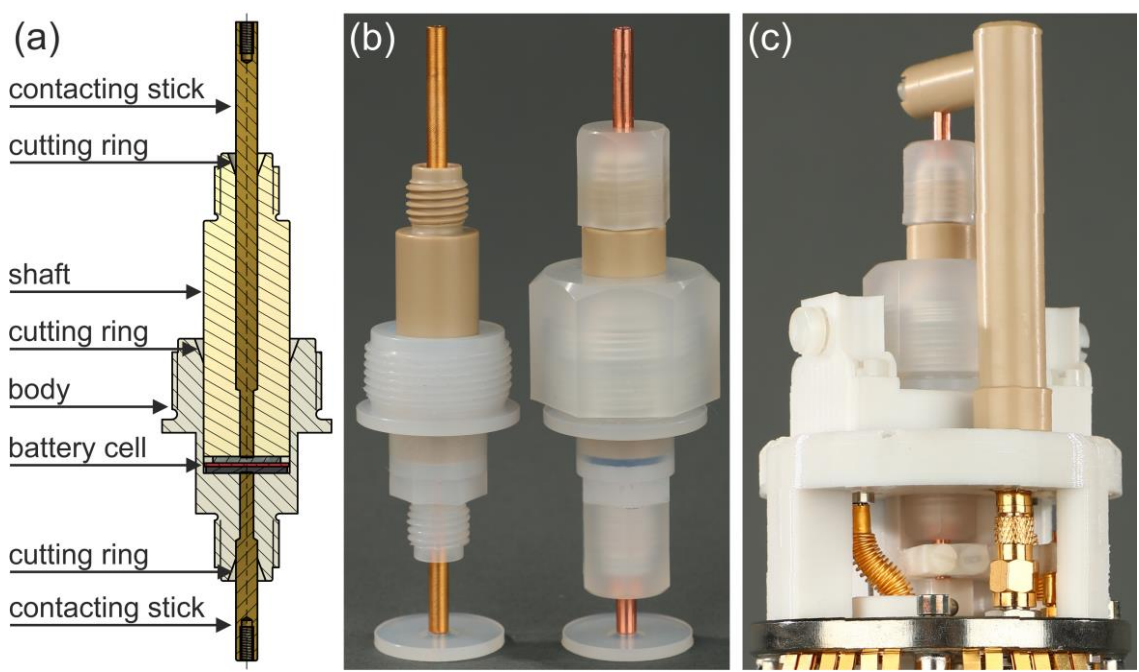

*Fig. 1: Design of in-operando NMR cell. (a) Schematic drawing and (b) photograph of cell container. (c) Connected cell mounted on the NMR probe. Battery electrodes are placed in a cylindrical body, which is closed with a shaft and sealed by a cutting ring. To connect the electrodes, plastic contacting sticks with conductive coating are pushed through holes in body and shaft. The sticks are sealed by cutting rings as well.*

The container was produced of PFA (Perfluoralkoxy, operation temp. up to 260 °C) and PEEK (Polyetheretherketone, operation temp. up to 280 °C). PFA is transparent, thus assembled components inside the body can be visually inspected (Fig. 1b). PEEK has a higher durability under mechanical stress[55–57] and is the material of choice for the contacting sticks. Most importantly, both materials have a high chemical resistance[58] not only against nearly all kinds of chemical substances commonly contained in battery electrolytes, but also against bases and acids as well as metallic lithium. The chemical resistance was tested by storing and weighting PFA, PEEK and PCTFE (Polychlorotrifluoroethylene, operation temp. up to 175 °C) disks in LP30 electrolyte for two years. PEEK lost its brownish color and became white, PFA and PCTFE were stable. By pressing Li-metal on PTFE and PCTFE, both became black. PFA and PEEK showed no changes after cleaning. Thus PFA for the body, shaft and cutting rings and PEEK for the contact sticks were used. For better visibility a PEEK shaft and PCTFE screw nuts were used for the photographs shown in Fig. 1.

### 2.2 RF Coil Simulations

The described cylindrical cell container is placed vertically inside the NMR magnet, thus oriented with its axis parallel to the $B_0$ field, and with the cell electrode plates horizontal, parallel to the $B_1$ field (Fig. 2). A saddle coil design is used for rf excitation, which was optimized numerically regarding its impedance, $B_1$ field strength and $B_1$ homogeneity. The simulations are based on a wire model of the coil design and performed using NEC4 ("Numerical Electromagnetics Code", Lawrence Livermore National Laboratory, Livermore, CA). The NEC software uses the method of moments, thus calculates the electromagnetic field for a setup that consists of a network of wire segments. As demonstrated by Dietrich and Sebak[35] we control the software by our own Matlab interface to adjust rf coil parameters and to plot the 3D simulation results. By specifying the wire material and wire radius, these simulations estimate the coil impedance. With this impedance value, we ensure that the coil can be tuned and matched in the resonance circuit of the probe. The 15 mm inner diameter of the modeled saddle coil is given by the cell container diameter. A height of 4 mm is chosen with respect to the size of battery cells. The wire diameter is 0.75 mm and its conductivity is set to that of copper. To account for disturbances from the contacting sticks of the in-operando NMR container, these cell connections are included in the model by adding passive vertical wires with a diameter of 2 mm along the center of the coil. These wires have a gap of 1.0 mm in the center of the coil where the battery electrodes are placed. The electrode plates itself are currently not included in the model.

With these simulations, the coil inductance and hence the impedance of the rf circuit could be adjusted close enough to a target value that tuning and matching could be performed using a set of variable capacitors built into the probe. To verify the accuracy of the simulation results, the spin nutation curve is estimated by using the simulated $B_1$ field and a $B_0$ field homogeneity of $\Delta B_0 = 6 \times 10^{-7}$ T. This value of $\Delta B_0$ is chosen with respect to a measured $^7$Li full width at half maximum (FWHM) of around 5 Hz at 9.39 T for 1 M LiCl solution, soaked into three layers of glass fibre separator (1 mm height) inside the cell chamber of the in-operando container. The calculated results are compared to measured nutation curves from the described setup. To investigate the effects of cell connections, simulations are compared with measurements performed with solid copper contacting sticks, pure PEEK sticks and PEEK sticks with a 500 nm gold coating. Gold was used instead of copper to ensure a high surface conductance in an aqueous solution of LiCl.

### 2.3. NMR Probe

The in-operando cell container is placed in a Bruker high-power broad-band high-temperature (HPBBHT) probe. The top of the probe was modified by attaching 3D-printed (Stratasys 3D-printer Fortus 400mc) parts for the container mounting and the saddle coil (Fig. 1c and label ① in Fig. 2). The middle of this coil is placed in the sweet-spot of a 400 MHz Bruker wide-bore NMR magnet. Using higher container mounting parts, the same probe was also tested in a 600 MHz Bruker wide-bore magnet. A broad-band frequency range of 55 MHz – 225 MHz becomes possible by an exchangeable resonant circuit capacitor.

The cell temperature can be controlled by placing a hood on top of the container mount and using a tempered nitrogen gas flow. Because polycarbonate 3D-printing was employed, the upper temperature limit is 120°C. The lower temperature limit depends on the cooling power of the gas flow and –30°C can be easily reached.

The electrical connections to the container contacting sticks consist of 50 μm copper wire (GVL Cryoengineering). Thereby metallic parts close to the rf coil and distortions of the $B_0$ field homogeneity are minimized. To avoid vibrations, the wire for the upper connection is embedded in a stable PEEK gallows (labeled with ② in Fig. 2) with SMB plug at the bottom. A small 3D-printed cable guide with attached SMB board connector is used for the 50 μm wire, which is pushed against the lower contacting stick (label ③ in Fig. 2a). Inside the probe frame, the inner conductors of two semi-rigid cables are used. At the bottom plate of the probe these cables are connected to low pass frequency filters (API Technologies Spectrum Control 51-311-316). Thereby a transmission of signals in the MHz frequency range via the battery cables towards the battery control device is reduced (70 dB @ 1 MHz to 1 GHz). Without low pass filters the rf pulses induce changes in the voltage measured by the potentiostat with peak heights up to 0.1 V. Because signals below 150 kHz are not strongly attenuated (<20 dB), electrical impedance spectroscopy (EIS) inside the NMR magnet can be performed as well. Battery testing and EIS experiments are performed using a Biologic SP-200 potentiostat/galvanostat.

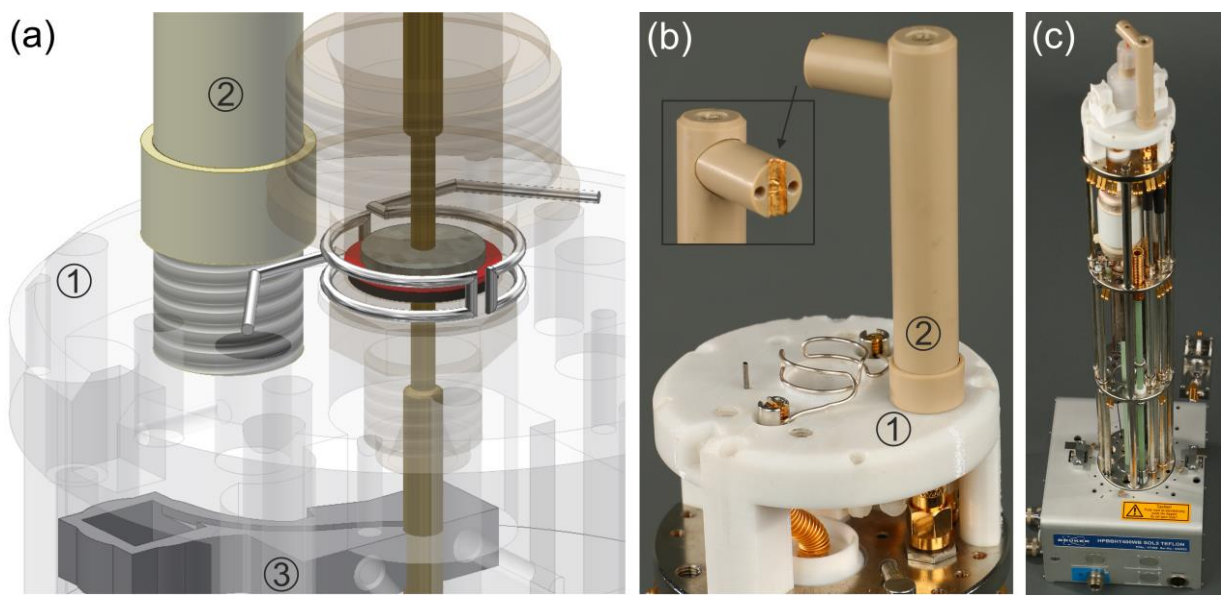

*Fig. 2:* *3D-printed mount on top of the NMR probe for in-operando cell container and rf coil. (a) CAD drawing of the 3D-printed mount (label ①) and the saddle coil around the cell container with battery electrodes and contacting sticks. For the connection of the cell a stable PEEK gallows (label ②) and a 3D-printed cable guide (label ③) are used. (b) Photography of the top of the probe with disassembled container mount. The coil geometry was optimized by numerical $B_1$ field simulations. (c) Photograph of the probe with connected cell.*

## 2.4. Battery Cell Materials and Experimental Parameters

Electrode materials (LCO and graphite) were purchased from Custom Cells Itzehoe GmbH. Sheets with 100 μm LCO on 20 μm Al foil exhibiting 3.5 mAh/cm$^2$ area capacity (155 mAh/g specific capacity) and 112 μm graphite on 18 μm copper foil exhibiting 3.8 mAh/cm$^2$ (350 mAh/g) were stamped to get 12 mm circular electrodes. In all experiments Whatman GF/D glass microfibre separators (approx. 0.1 mm to 0.3 mm thickness after compression) and LP30 electrolyte (1 M LiPF$_6$ in 1:1 ethylene carbonate:dimethyl carbonate) from BASF were used. For the current rate capability comparisons with LCO/Li-metal batteries one separator and battery grade metallic Li from Rockwood Lithium GmbH (now Albemarle) with a thickness of 250 µm were used. The standard cell[59] used as reference was assembled in a Swagelok PFA housing with stainless steel current collectors, a spring and a 12 mm nickel disk on top of the Li-metal. The Li-metal/graphite battery for in-operando NMR experiments was assembled with Rockwood Li-metal, Custom Cells graphite and three separators.

In-operando NMR measurements were performed in a 9.39 T Bruker Ascend 400WB spectrometer system. The HPBBHT probe temperature was stabilized to 25 °C with 800 l/h nitrogen gas flow. For excitation 500 W pulse power, 6 µs pulse length and a scan delay of 5 s were used.

For winding a 0.75 mm silver-coated copper wire to a numerically calculated saddle coil shape, a 3D-printed cylinder template was used. By shimming the $B_0$ field of the magnet with a battery container a minimum $^7$Li FWHM of approximately 5 Hz was reached. This was achieved both for contacting sticks completely made of copper and for plastic sticks with thin-film contacting layers. For this experiment and for spin nutation curves three glass fibre separators on top of each other, with a total height of 1 mm in the 12.5 mm inner diameter cell chamber, were filled with 100 µl of 1 M LiCl in water.

For EIS measurements a Li-metal/graphite battery was cycled between 1 mV and 3 V with stops (hold for 20 minutes) at 150 mV during charge and 165 mV during discharge. Before the EIS measurement, the system was allowed to equilibrate for 10 minutes. The impedance was measured with a potentiostat with built-in frequency response analyzer (Bio-Logic VSP-300). Measurements were conducted from $f = 2 \times 10^5$ Hz to $f = 1 \times 10^{-2}$ Hz with an excitation amplitude of $U = 1 \times 10^{-2}$ V. The temperature was kept constant at 30 °C in a climate chamber (Binder KB115). For a higher resolution, the impedance results are transformed into the relaxation time domain[60].

Peak fitting to separate the signals of metallic, mossy and dendritic structures was performed with Matlab. Three mixed Gaussian - Lorentzian lines with position and width confinements were used. For the peak center positions a variation of 5 ppm was allowed and the peak widths were confined between 5 ppm and 30 ppm.

**2.5 Signal referencing in NMR data analysis**

Quantifying the evolution of Li in the different Li containing components in the cell is not straightforward. For one, electrically conducting metallic components shield electromagnetic radiation (skin effect). Only if the skin depth $\delta$ is much larger than the thickness $d$ of a metallic component, then the corresponding NMR signal is proportional to the number of nuclear spins in the sample. Furthermore, certain parts of a sample may be shielded by other parts, depending on the sample geometry and the direction of the $B_1$ field. Eventually, if Li ions are transported between different environments, such as a metallic electrode and an intercalation compound, the rf impedance of the sample can change. This affects tuning and matching of the NMR resonant circuit and hence its quality factor, leading to a changing signal scaling. While an in-depth discussion is outside the scope of this work, a simple and robust approach for a qualitative discussion is represented by the use of the Li-metal signal $I_{LM,0}$ of the pristine cell as reference. Since there is a large surplus of Li in the Li metal electrode compared to the capacity of the graphite electrode, the thickness $d_{LM}$ of the Li metal electrode is always larger than the skin depth $\delta_{LM}$. Therefore the variation of $I_{LM}(t)$ represents a rough estimate for the accuracy of this referencing method.

To quantify signal changes, $dI/dt$, an arbitrary scaling factor for the slope of $\Delta I_{LM,0} = I_{LM,0} / 1h$ was chosen, i.e. how much a particular signal changes in one hour compared to the amplitude of the metallic Li signal of the pristine cell.

## 3. RESULTS AND DISCUSSION

### 3.1 Current rate capability and long-time cycling

The charge–discharge capacities with current rates of C/10, C/5, C/2 and 1C (= 3.5 mA/cm$^2$) were measured for LCO vs. Li-metal cells in an in-operando container with thin-film contacting sticks and, as a reference, in a standard Swagelok PFA housing (Fig. 3). The utilized LCO electrodes enabled a maximum current of 1C. Differences of the cell capacities for both cells became more pronounced with increasing current rate. While the standard cell was assembled with a spring, no spring was used for the in-operando container because this can disturb NMR measurements.

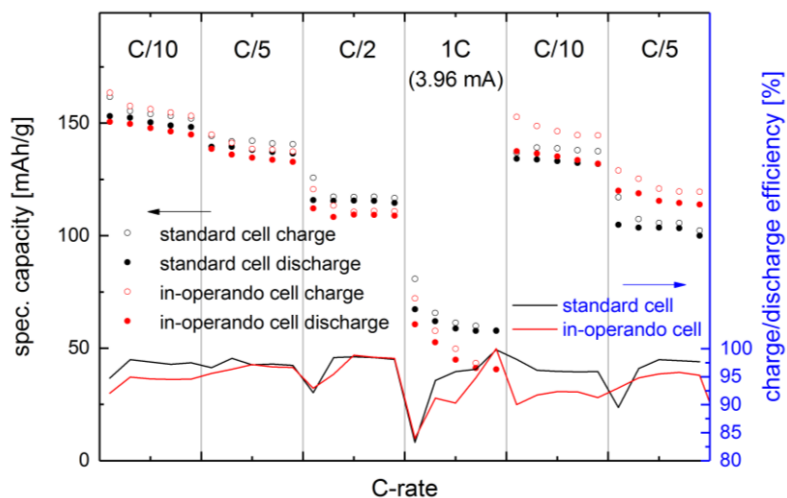

*Fig. 3: Comparison of the current rate capabilities of LCO vs. Li-metal cells in an in-operando container with 500 nm contacting layers (red) and in a standard Swagelok cell (black). The cells were cycled in a voltage range from 3.0 V to 4.3 V. Dots indicate the specific capacity for charging (open) and discharging (filled). Lines give the charge/discharge efficiency.*

The correspondingly larger pressure applied on the standard cell could be the reason for its higher rate capability. For the five repetition cycle with C/5 the in-operando cell exhibited around 15 mAh/g higher capacities but smaller charge/discharge efficiencies. This efficiency is given by the ratio of discharge capacity divided by charge capacity. For both cells it was above 90 %, except for the first cycle with 1C current. The comparison of the current rate capabilities indicated a suitable cycling behavior of the in-operando cell container.

To show the long-time reliability of the developed in-operando NMR container with thin-film contacting sticks, a LCO vs. graphite battery cell with one glass fibre separator was charged and discharged galvanostatically for 2400 h. The cell capacity degradation and voltage profile are shown in Fig. 4a. After the first cycles for formation a current of C/4 (1 mA), calculated according to the theoretical capacity of the cell, was applied for cycling the cell until the discharge capacity dropped to zero. After approximately 1040 h and 340 cycles the remaining discharge capacity was just above 0.1 mAh. The cell was then refreshed by applying currents of C/40 (0.1 mA) for one cycle, followed by C/16 (0.25 mA) for two cycles. Hereon the cycling was continued with C/8 current in the voltage range from 3.3 V to 4.5 V for approximately another 750 h and 240 cycles. Finally the current was reduced to C/16 and 200 additional cycles were performed between 3.0 V and 4.5 V. Thereby a battery operation of more than 2400 h in total was accomplished in the in-operando container. The most obvious reason for the degradation of the battery is overcharging[19,61,62] caused by the high voltage limit of initially 4.3 V and later 4.5 V. After 2000 h the cell still exhibits a capacity of around 0.5 mAh. For comparison 12 mm electrodes with high power materials have an initial cell capacity of around 0.5 mAh as well[61].

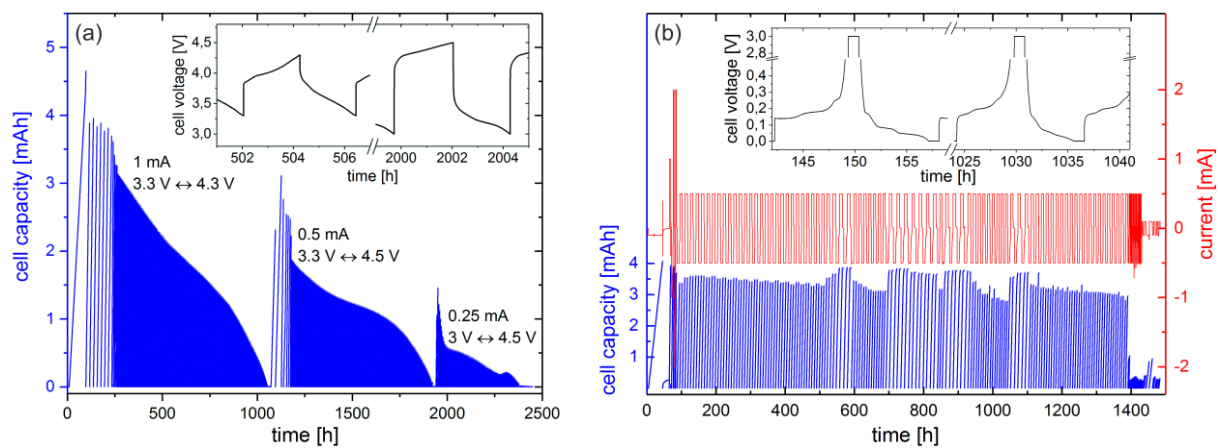

*Fig. 4:* Cell capacity degradations of a LCO vs. graphite and a Li-metal vs. graphite in-operando battery. (a) Cycling of LCO vs. graphite battery during 2400 h. For the first four cycles within 230 h a voltage range of 3.0 V to 4.3 V and low currents (C/80 (0.05 mA) and C/20 (0.2 mA)) were used. Then the cut-off voltage was set to 3.3 V and the current was increased to C/8 (0.5 mA) for one cycle. The next cycle was performed with C/4 (1 mA) for charging and C/8 for discharging. After the first cycles a current of C/4 (1 mA) was applied. The cell was refreshed with C/40 and C/16 current after a capacity drop to zero. Then the current was set to C/8 and later to C/16. The cell lifetime even over long times seems to be limited by materials and protocols, not cell housing. (b) Cell capacity degradation (blue), charge/discharge current (red) and voltage profile (insert, black) of a Li-metal vs. graphite battery with three separators assembled for long-run in-operando NMR measurements. The voltage range was 3 V to 1 mV. After the first discharge with C/43 (0.1 mA) and a voltage-hold at 1 mV for 20 h, one cycle was performed with C/4.3 (1 mA) followed by 1.5 cycles with C/2.15 (2 mA). Then currents of C/8.6 (0.5 mA) were applied. Different voltage-hold times of 1 h to 4 h were tested.

This demonstrates that properties of the employed material in combination with the cycling program used in the experiment was limiting the cell lifetime, while the contribution of the in-operando container was not considerable.

For a Li-metal vs. graphite battery with three separators slower capacity degradation was found (Fig. 4b). Even after 1390 h cycling with C/8.6 (0.5 mA) the cell still exhibited a capacity of almost 3 mAh (2.65 mAh/cm$^2$). Three separators were chosen to enable a long lifetime despite dendrite growth, which led to a short circuit after 88 cycles within 1400 h (Fig. 5). Thereby long-run in-operando NMR measurements of this cell became possible. The NMR results are shown in chapter 3.3.

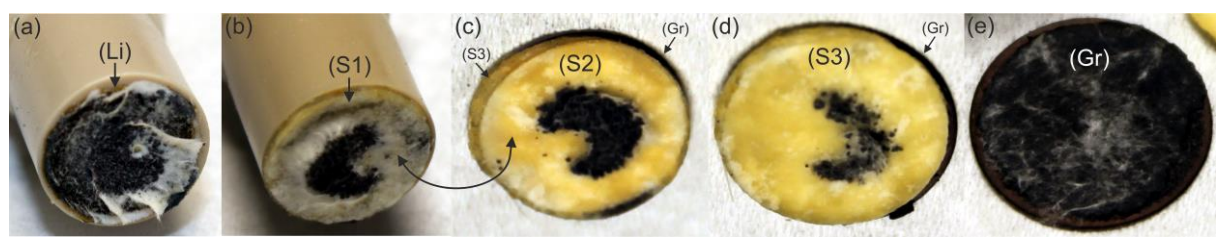

*Fig. 5:* Post-mortem photographs of the Li-metal vs. graphite cell taken during disassembling. Li microstructure formation through three separator layers led to a short circuit. (a) Li-metal electrode (Li) on shaft. (b) Separator 1 (S1) on Li-metal electrode. (c) Separator 2 (S2) on separator 3 (S3) on graphite electrode (Gr). Before lifting S2, S3 and Gr from separator 1 on the Li-metal electrode, the two visible faces of S1 and S2 have been in contact. (d) Separator 3 on graphite electrode. (e) Graphite electrode.

## 3.2 EIS of an in-operando Li-metal vs. graphite battery

After the described in-operando NMR measurements of the first three and a half cycles (chapter 3.4), in-situ electrical impedance spectroscopy (EIS) measurements were performed during the following eleven cycles. For this, the cell voltage was kept constant on plateaus at 150 mV during charging (25 % state-of-charge (SOC)) and at 165 mV during discharging (85 % SOC). The distribution of relaxation times (DRT) of the EIS measurements are shown in Fig. 6a. The red curves show the DRT spectrum during charging and the blue curves during discharging. Four peaks, caused by different processes with different time constants, were identified and quantified by peak-fitting. The time constants $\tau$ of the individual processes, given by the polarization resistance $R$ times the polarization capacitance $C$, are depicted in Fig. 6b. The fastest two processes starting with $\tau \approx 3$ µs (blue) and $\tau \approx 50$ µs (red) were most likely caused by electronic interactions at the electrodes and the separator[63,64]. The process with $\tau \approx 0.3$ ms (green) could be attributed to the SEI layer[65]. The slowest process with $\tau \approx 2$ ms (yellow) was most likely caused by ionic charge transfer in the active materials[66,67].

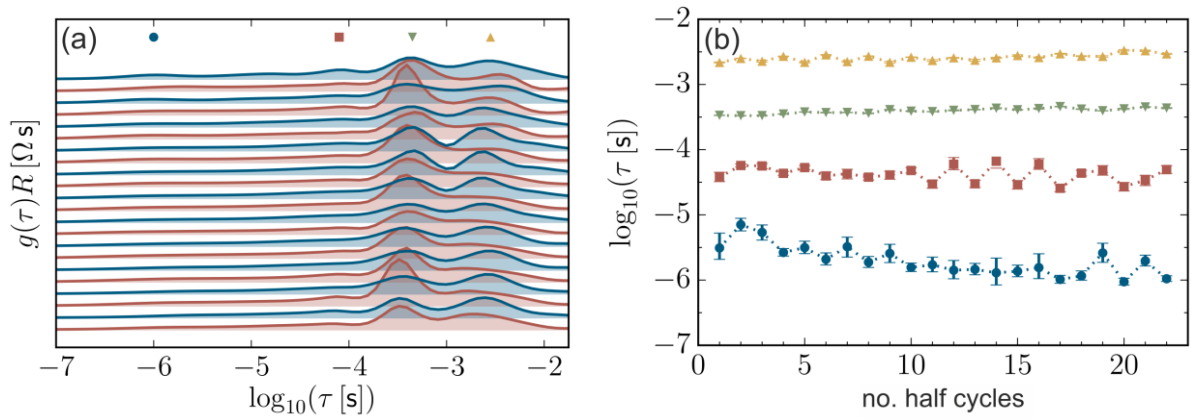

*Fig. 6: Electrical impedance spectroscopy (EIS) of Li-metal vs. graphite cell. (a) Relaxation time spectra of impedance data measured at consecutive half-cycles at 25 % SOC during charging (red) and at 85 % SOC during discharging (blue). $g(\tau)$ is the spectral density, representing the distribution of relaxation times $\tau$, and R is the polarization resistance. Four peaks, caused by different processes with different time constants, could be identified. (b) Evolution of the time constants determined by peak fitting. The different processes are assigned to electronic interactions between electrodes and separator (blue, red), the SEI layer (green) and charge transfer impedance (yellow).*

An increase of the SEI time constant from $\tau \approx 0.3$ ms to $\tau \approx 0.5$ ms indicated, in accordance with literature, changes and reformation of the SEI layer[49,68]. Because the thickness of an SEI layer can be well below 100 nm, NMR is not sensitive enough to follow the initial SEI formation. EIS measurements are a viable option to close this gap. In case of an NMR investigation of the electrolyte stability, a correlation between electrolyte decomposition and SEI formation is necessary.

While on the long run the time constant of the fastest electronic interaction process decreased from $\tau \approx 3$ µs to $\tau \approx 1$ µs, the ionic charge transfer impedance slowly increased from $\tau \approx 1$ ms to $\tau \approx 3$ ms. This could indicate an increase of the electronic conductivity of the graphite electrode, which could be caused by lithiation[69]. In agreement with the NMR data, the decrease of the ionic conductivity could be caused by the growing 'dead' Li microstructure layer.

### 3.3 Simulations and measurements of spin nutation curves

To test the $B_1$ magnetic field strength and homogeneity of our custom 15 mm saddle coil with solid copper contacting sticks, PEEK sticks with 500 nm gold coating and non-conductive PEEK sticks, spin nutation curves, using the in-operando NMR container with LiCl solution, were acquired. On-resonant FIDs were measured with 200 W rf pulse power for pulse durations varied between 0 μs and 600 μs in 2.5 μs steps (Fig. 7d). These nutation experiments were compared with simulated nutation curves (Fig. 7c) based on the calculated $B_1$ field distribution of the rf coil (Fig. 7a,b). Similar ratios of the intensity of the $N^{th}$ maximum divided by the intensity of the first maximum were observed. The differences of the nutation curves with and without conductive sticks were small. Thin-film contacting layer were not included in simulations so far. In experiments an influence of the contacting sticks was not observed for pulse lengths below the π-pulse length of 12.5 μs. Close to the first intensity minimum at around 80 μs, minor distortions of the nutation curve were measured reproducibly with solid copper sticks. For pulses longer than approximately 180 μs all three nutation curves differed slightly. The positions of the maxima were similar for pure PEEK and PEEK with 500 nm coating. For solid copper sticks a slightly faster oscillation was found.

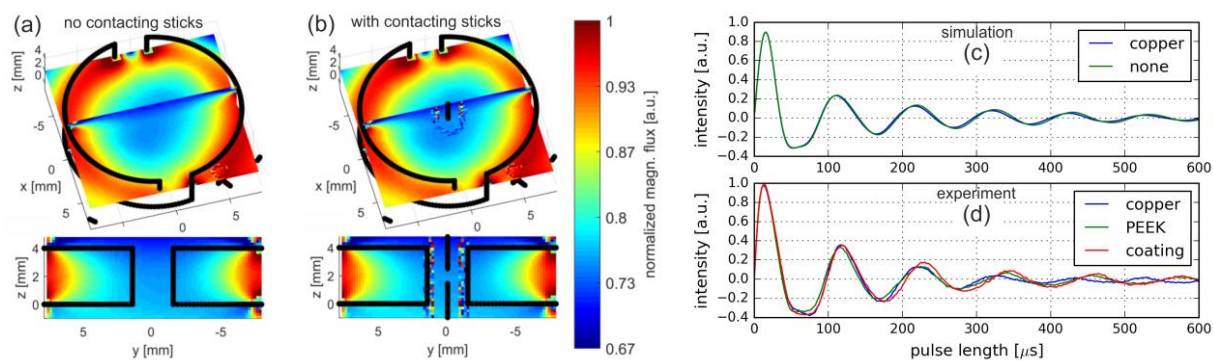

***Fig. 7:*** *Numerical simulations of the $B_1$ magnetic field strength and comparison of calculated and measured nutation curves for different contacting sticks. (a, b) Black lines indicate the position of the rf saddle coil and the contacting sticks (only b) in the center. Two slices are shown for $B_1$ field simulations (a) without contacting sticks and (b) with solid copper contacting sticks. (c) From $B_1$ field simulations calculated nutation curves with and without copper contacting sticks. (d) Measured $^7Li$ signal intensity as a function of rf pulse length for solid copper, non-conductive PEEK as well as PEEK contacting sticks with 500 nm gold coating.*

### 3.4 Long-run in-operando NMR of the lithium microstructure formation

A Li-metal vs. graphite battery with three glas fiber separators and a cell thickness of 0.3 mm was investigated by $^7Li$ in-operando NMR. Fig. 8a shows the charge/discharge profile of the first three cycles with a current of C/10 (0.37 mA cm$^{-1}$) after an initial open circuit voltage (OCV) period of 10 h, and Figs. 8b and 8c show the time-resolved NMR spectra. The spectra were processed using an average over 80 scans with a scan delay of 5 s and the integrals shown in Fig. 8d were obtained by peak fitting. Signals of the LiPF$_6$ and its products in the electrolyte were composed of at least two overlapping resonances between –1 ppm and 2 ppm[18,70]. For the most intensive signal of mobile Li$^+$ ions at –0.5 ppm a FWHM of about 150 Hz (less than 1 ppm) was observed. The Li$^+$ peak intensity divided by the noise standard deviation of the processed spectra gave a signal-to-noise ratio of approximately 2500. The evolution of the electrolyte Li$^+$ signal integral is shown in Fig. 8d (black line). Most likely the Li distribution in the electrolyte was dependent on the applied current and changed during cycling[41]. An increase of the Li$^+$ concentration in regions with high $B_1$ field strength could lead to an enhanced signal. In addition to the electrolyte peaks, the initial NMR spectra showed a single signal of the Li-metal electrode with a FWHM of approximately 1.1 kHz (7 ppm) centered at 244 ppm and with a signal-to-noise ratio of approximately 500.

During OCV at a state-of-charge (SOC) of 100%, the Li-metal and $Li^+$ signals were stable. Shortly after the beginning of discharging (at a SOC of about 93%), the peak centers of both signals shifted by approximately 300 Hz (2 ppm). During lithiation of the graphite electrode the transition of diamagnetic to paramagnetic graphite takes place[71]. The corresponding susceptibility changes could be the reason for the center frequency shifts. By reversing the current from discharge to charge no similar effect was observed. However, at the end of delithiation, again at an SOC of about 93%, the peaks shifted back to their initial frequency positions. The line shapes were not affected. In addition to these shifts, the graphite lithiation led to the growth of a signal centered at approximately 13 ppm at the beginning of discharging. This was reported to be originating from $LiC_{18}$ as a result of Li intercalation[46]. Then signals for $LiC_{12}$ and $LiC_6$ grew at 43 ppm and 38 ppm, respectively, accompanied by a decrease of $LiC_{18}$. The reverse process occurred during charging and the NMR spectra showed the delithiation of the graphite electrode. The corresponding voltage plateaus were in accordance with the known staging behavior in graphite[72].

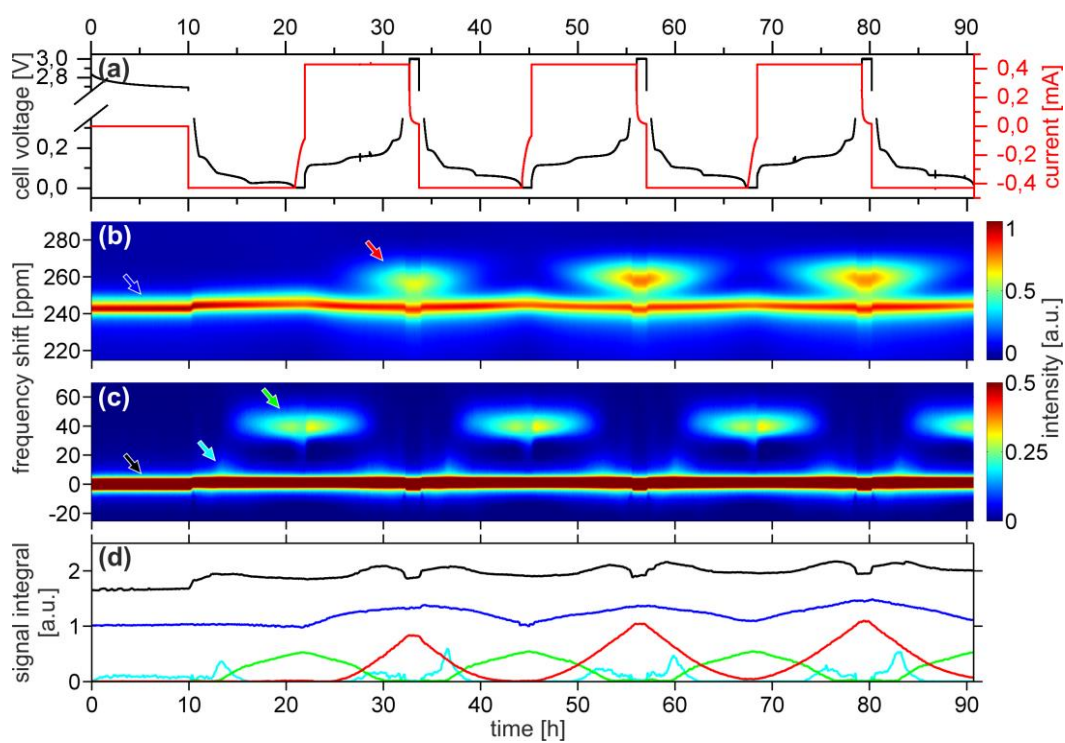

***Fig. 8:*** *In-operando $^7Li$ NMR spectra of the first three and a half cycles of a Li-metal vs. graphite battery. (a) Charge/discharge voltage and current profiles. (b) Pseudocolor plot of the Li-metal (244 ppm, blue arrow) and Li microstructure (258 ppm, red arrow) signals. The signal was normalized for an initial Li-metal peak intensity of 1. (c) Pseudocolor plot of the electrolyte (-0.5 ppm, black arrow), $LiC_{18}$ (13 ppm, cyan arrow), $LiC_{12}$ (43 ppm) and $LiC_6$ (38 ppm, green arrow) signals. (d) Signal integral evolution of Li-metal (blue), microstructured Li (red), $Li^+$ in electrolyte (black), $LiC_{18}$ (cyan, magnified by factor 5), and $LiC_{12}$ combined with $LiC_6$ (green). The microstructure formation on the Li-metal electrode started with the first charging. By discharging these Li structures were consumed again. The integrals were rescaled such that the initial Li-metal signal integral $I_{LM,0}$ was equal to 1.*

Changes of the Li-metal surface morphology during the first discharge were indicated by rf phase changes of the Li-metal signal in combination with a broadening towards lower field by approximately 300 Hz and a slow decrease of the Li-metal signal integral $I_{LM}$ by around 3%. The reason could be a roughening of the Li-metal surface[49,73]. Charging the cell for the first time led to an increasing Li-metal integral $I_{LM}$ by approximately 35% and a Li microstructure signal with a FWHM of about 2.3 kHz (15 ppm) centered at approximately 258 ppm. The increase of the Li-metal signal is in accordance with experiments and simulations in which skin depth issues have been taken into account[12]. A microstructure formation during the first Li deposition on the Li-metal electrode is consistent with several SEM[50,73], optical microscopy[74], X-ray tomography[75] and electron paramagnetic resonance (EPR)[44] investigations. The in-operando NMR data showed an almost linear growth and decrease of the Li microstructure signal integral $I_{\mu L}$ for the second half of charging and the corresponding first half of discharging, respectively, as expected for a constant current experiment[12,29,36]. With one electron at most one Li$^+$ ion can be reduced and deposited on the Li-metal electrode and vice versa[42].

A slope of $dI_{\mu L}/dt \approx 0.13\ \Delta I_{LM,0}$ was calculated by a linear regression during charge, and during discharge the Li microstructure signal decreased to almost zero with approximately the same but negative slope of $dI_{\mu L}/dt \approx -0.13\ I_{LM,0}$ h$^{-1}$. This indicated that the previously formed Li microstructure was reversibly removed. Nevertheless, the Li microstructure NMR signals only vanished completely after the first discharge. Remaining signals were found after the next two discharges in combination with an increased maximum Li microstructure signal at the end of each charging. This is consistent with optical microscopy investigations, which revealed the formation of Li microstructures growing out of holes and thereby pushing away the 'old' microstructure layer[74]. Vertically growing microstructures can percolate the battery separator, as shown in Fig. 5. Once broken or oxidized far enough, they can lose contact, thereby get electrochemically inactive and thus cannot be removed[76]. On the long run these porous Li structures can become large enough to cause a short circuit between the electrodes.

To further investigate the formation of microstructured Li, a different Li-metal vs. graphite battery was cycled for around 1400 h (Fig. 4b) with C/8.6 (0.5 mA). When assembling, the electrodes and the three glass fiber separators were compressed less strongly than in the first container until a cell thickness of 0.5 mm was obtained. Due to the duration of the experiment, NMR measurements could only be conducted stroboscopically. The evolution of the Li-microstructure signals is shown in Fig. 9. The battery operation was started with a long discharge for 60 h with C/43 (0.1 mA), including 20 h voltage hold at 1 mV to search for Li microstructure signals caused by overlithiation and plating of the graphite electrode[19]. A Li microstructure signal increase of slightly less than 5% compared to the pristine Li-metal signal $I_{LM,0}$ was found. At the same time, a shifting Li-metal rf phase and a decrease of the signal integral $I_{LM}$ by approximately 5% was observed, indicating changes of the Li-metal surface morphology occurring already during the first discharge. Alternatively, a small amount of metallic Li on top of the graphite electrode may have formed during discharging[19]. Within the first three cycles, $I_{\mu L}$ reached an approximately twice as high maximum value than in the first cell. This is consistent with the finding that pressure on the electrodes reduces the microstructure layer thickness[16]. Another cause could be the increased current rates during the first couple of cycles.

Using a C/8.6 current for long-run operation, with each cycle both the maximum microstructure signal after charging and the minimum signal after discharging increased. Microstructure layers are poorly connected to the Li-metal surface via a few contacting points that can be dissolved during discharging[44,50,77]. As a consequence, parts of the microstructure can become electrochemically inactive and thereby cannot be removed anymore. These disconnected structures, called 'dead Li', appear to grow quite continuously, which indicates that for a given C-rate in a particular cell the same amount of Li is irreversibly deposited per cycle[44]. For the C/8.6 cycles a linear fit was used to approximate the increase of residual microstructure. A slope of 0.04 $\Delta I_{LM,0}$ for both the maxima and minima of the long-term evolution of $I_{\mu L}$ was found (yellow and green lines, respectively, in Fig. 9b). Different voltage hold times most probably led to small deviations from an exactly linear growth, but non-uniform growth of microstructured Li (Fig. 5) may be another factor.

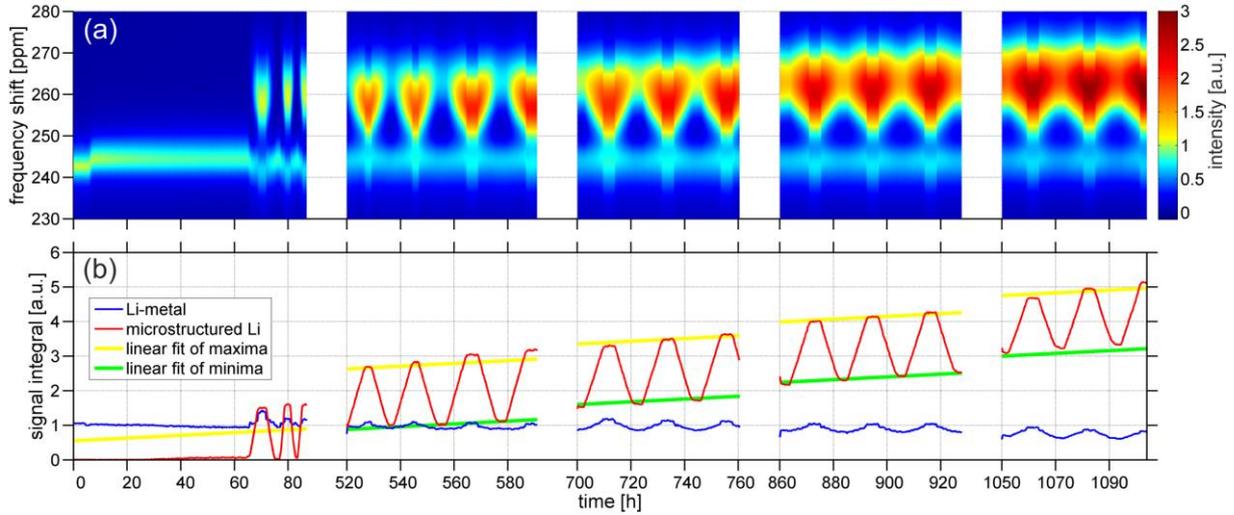

*Fig. 9:* Long-run in-operando NMR measurements of the Li microstructure formation in a Li-metal vs. graphite battery. (a) Pseudocolor plot of the Li-metal (244 ppm) and Li microstructure (260 ppm) signals. Within 1100 h the microstructure center frequency shifted by around 600 Hz toward lower field and the signal reached an intensity of more than five times the pristine Li-metal signal. (b) Signal integral evolution of Li-metal (blue) and microstructured Li (red). A linear increase of the maxima and minima of the microstructure signal was approximated by to lines (yellow and green) with a slope of 0.04 $\Delta I_{LM,0}$. $I_{LM,0}$ in (a) was normalized to 1. The integrals in (b) were rescaled such that the signal integral of $I_{LM,0}$ was equal to 1. The charge/discharge profile is depicted in Fig. 4b and post-mortem photographs are shown in Fig.5.

The line approximating the minima evolution could be extrapolated to the signal integral $I_{\mu L,min}$ at the point of the short circuit at $t=1400$ h, yielding a value of $I_{\mu L,min}(t=1400$ h$) = 4.4$ $I_{LM,0}$. It is tempting to speculate that long-run in-operando studies could give an estimate for the Li microstructure signal $I_{\mu L}$ at which a short circuit can be expected and use this to estimate the state-of-health at a particular point in time or the cell lifetime under different operating conditions by extrapolation from a much shorter run. However from Fig. 5 it is evident that the formation of dead Li is not necessarily uniform in the whole cell, therefore the suitability of such a method requires better statistical evidence from a larger set of identical test cells.

After an operation time of 1100 h the microstructure signal at its maximum within a cycle reached a value of more than five times the pristine Li-metal signal integral $I_{LM,0}$. Since for thick layers of bulk Li metal the signal is eventually limited by the skin effect, such a signal increase indicates a fairly complex dependence of the signal amplitude on the cell geometry as well as the morphology of the Li metal. This shows that each signal component in the Li NMR spectrum of a battery cell needs to be referenced independently for a quantitative data analysis.

Cycle by cycle the average center frequency of the microstructure signal shifted toward lower field (Fig. 9a). Compared to the first cycle, a shift of about 600 Hz (4 ppm) was measured after 1100 h. Slow irreversible growth of porous metallic lithium into the previously diamagnetic electrolyte may increase the effect of Knight shift and thereby cause such a frequency shifts toward lower field[19]. This long-term shift was accompanied by a shift of the microstructure frequency within each cycle by about 3 ppm. In our cell we did not observe an alternating growth and decline of two separate resonances, but a fairly continuous state-of-charge dependent shift, which may be caused by a gradual morphology or density change of the microstructured Li between charged and discharged state of the cell[16].

Furthermore, the almost instantaneous shift by about 1.5 ppm once a certain degree of graphite intercalation was reached, which was previously described for the first cell, could be observed as well for the full cell lifetime. While the resonance frequency of the microstructured Li was fairly selective regarding the state of the cell, the FWHM only changed little between about 15 ppm for the fully charged cell and about 13.5 ppm (2.1 kHz) for the discharged cell.

The slope of the microstructure signal during cycling was showing a somewhat different behavior than for the first cell that was more strongly compressed during assembly. First of all, $I_{\mu L}$ increased and decreased fairly linearly while Li was deintercalated and intercalated, respectively, from/into the graphite. Using C/8.6 current, the signal evolution after 520 h showed an average slope of 0.27 $\Delta I_{LM,0}$ during charge and of -0.27 $\Delta I_{LM,0}$ during discharge. After 1050 h an average slope of 0.23 $\Delta I_{LM,0}$ during charge and of -0.23 $\Delta I_{LM,0}$ during discharge was observed. For the C/2.15 (2 mA) and the C/4.3 (1 mA) cycles immediately after the formation cycle the signal evolution was less linear, which may have been caused by a less uniform microstructure formation and increased consumption of the bulk metallic Li electrode due to the increased current density[76–79]. The slope of $I_{\mu L}$ increased for C/2.15 during charging up to 0.8 $\Delta I_{LM,0}$ and decreased during discharging down to at most -0.7 $\Delta I_{LM,0}$. For C/4.3 a maximum signal increase rate of 0.55 $\Delta I_{LM,0}$ and a maximum decrease rate of -0.5 $\Delta I_{LM,0}$ were found. The difference of the slope between charge and discharge, while at the same time the cell efficiency stayed fairly high (Fig. 4b), points towards an increased rate at which the bulk Li was consumed at higher C-rates.

## 4. CONCLUSIONS

For long-run in-operando NMR measurements a cylindrical NMR battery container was developed. A reliable cycling behavior and the long-time gas-tight sealing of the container design were demonstrated by current rate capability tests and charge–discharge cycles for more than 2400 h. Plastic sticks with thin-film conductive coatings were used for contacting the electrodes. Thereby the amount of metal inside the cell container is minimized and a variety of electrode connection designs become possible. Moreover it enables reducing disturbances of the magnetic field, screening of the NMR pulses and minimizing artifacts in imaging applications. To further decrease the thickness of conductive parts in the cell container, plastic carriers for the electrode materials with thin-film conductive coatings could be used instead of attaching the electrode materials on copper or aluminum foils.

The container was placed in a custom 3D-printed mount with a numerically optimized saddle coil, fixed on a commercial NMR probe. A comparison of numerically estimated with experimental spin nutation curves demonstrated a good agreement of experiment and simulation.

The formation of microstructured Li on a Li metal electrode was investigated. Gradual morphology or density change of the microstructured Li between charged and discharged state of the cell most likely caused a state-of-charge dependent frequency shift[16]. Even after 1000 h a quite continuous growth of 'dead Li' was measured, which indicates that the same amount of Li is irreversibly deposited per cycle for a given C-rate in a particular cell[44]. These findings indicate that when using a long-term stable cell, in-operando NMR is a suitable tool to characterize the complex dependence of the microstructure morphology, the thickness of the porous Li layer as well as enhanced current densities on the battery performance and aging[76–80]. This might help to identify successful dendrite mitigation strategies, such as enhanced cell pressure, additives or different load profiles. However, since dendrite formation is not necessarily uniform inside a cell (as seen in Fig. 5), further investigations are necessary to assess whether it is possible to predict the state-of-health or the lifetime of a cell by extrapolation from a short in-operando NMR run. In addition, non-uniform dendrite formation has the effect that the presented approach is only semi-quantitative, since the skin effect may lead to an individual variation of the $B_1$ distributions inside each cell. Also, a changing Li morphology during cycling induces temporally changing shielding of certain parts of the cell and could alter its rf impedance, which would influence the signal amplitude. Hence careful referencing is indispensable for quantitative applications.

## 5. ACKNOWLEDGEMENT

Thanks to Sven Reiter, Heinrich Pier and Knut Dahlhoff from ZEA-1 of Forschungszentrum Jülich for the container manufacturing, to Norbert Merl and Torge Thönnessen from Custom Cells for discussions and supply with electrode materials and to Helmut Timmermanns from ZEA-2 of Forschungszentrum Jülich for the photographs of the setup. Financial support by the German Federal Ministry of Education and Research (BMBF; Project AlSiBat) is gratefully acknowledged.